\documentclass[aps,showpacs,pra,twocolumn] {revtex4}

\usepackage{amsmath }
\usepackage{amssymb}
\usepackage{epsfig}
\usepackage{bm}

\input colordvi

\def\be{\begin{equation}}
\def\ee{\end{equation}}
\def\bea{\begin{eqnarray}}
\def\eea{\end{eqnarray}}

\def\ptl{\partial}

\begin{document}
\bibliographystyle{revtex}

\title{ Can the scale factor be rippled? }

\author{S.L. Cherkas}
\email{cherkas@inp.bsu.by}
 \affiliation{Institute for Nuclear Problems, Bobruiskaya
11, Minsk 220050, Belarus}
\author{V.L. Kalashnikov}
\email{v.kalashnikov@tuwien.ac.at}
 \affiliation{Institut f\"{u}r Photonik, Technische
Universit\"{a}t Wien, Gusshausstrasse 27/387, Vienna A-1040,
Austria}

\received{ \today }

\begin{abstract}We address an issue: would the cosmological scale factor be a locally
oscillating quantity? This problem is examined in the framework of
two classical 1+1-dimensional models: the first one is a string
against a curved background, and the second one is an inhomogeneous
Bianchi I model. For the string model, it is shown that there
exist the gauge and the initial condition providing an oscillation
of scale factor against a slowly evolving background, which is not
affected by such an oscillation ``at the mean''. For the
inhomogeneous Bianchi I model with the conformal time gauge, an
initially homogeneous scale factor can become inhomogeneous and
undergo the nonlinear oscillations. As is shown these
nonlinear oscillations can be treated as a nonlinear gauge wave.
\end{abstract}
\pacs{ 11.25.-w, 04.20.Ex, 98.80.-k } \maketitle

\section{Introduction}

Cosmological constant problem is a subject of numerous
investigations \cite{weinberg,carr,pd,ell,stein,arm,rev}. The fact of the matter is that the enormous energy of zero point quantum
fluctuations of matter fields as well as electromagnetic and
gravitational waves have to produce an enormous universe expansion,
which is not seen in reality. At the same time, a study of conformal fluctuations also has a long history
\cite{pad,mot,maz,cun} and is still far from the completion
owing to lack of a developed theory of quantum gravity.

A possible relation between these two problems has been considered in
Refs. \cite{wang,cong}.  It has been suggested that, since the
conformal fluctuations formally give a negative contribution to
the Hamiltonian density, they can compensate an energy of the zero
point fluctuations of matter fields, electromagnetic and
gravitational waves. Nevertheless, as was shown in \cite{bon}, it is not the case, at least
for the gravitational waves. The reason is that the scale (conformal) factor is attributed naturally to the slowly varying
background, and thus, it can hardly be modulated by
gravitational waves.

It should be added, that the decomposition of a metric into the conformal
factor and a remaining part: $g_{\mu\nu}=a^2\tilde
g_{\mu\nu}=e^{2\alpha}\tilde g_{\mu\nu}$ seems very natural as it
was suggested by York \cite{York}, who discussed properties of the
conformal geometry given by $\tilde g_{\mu\nu}$.

In this paper we investigate the classical solutions of  some
simple models to determine a possibility of the scale factor
modulation. This means that one explores some particular gauges
and initial conditions, which result in an oscillating scale
factor. This does not mean that the scale factor is an oscillating
quantity under an arbitrary initial conditions. The aim of our
work is to demonstrate, that the scale factor oscillations are not
forbidden in general (e.g. owing to constraints).

Although, we investigate local oscillations of the scale factor
here, the models admitting global oscillations \cite{bar1, bar2,
bar3, ah} are also of general interest.

\section{String in a background space}

The first model considered is
a string against a curved background. For this model we have built
a solution for which the scale factor is connected rigidly with
the matter fields so that a mutual compensation of their
fluctuations and, thereby, the solution of the cosmological constant problem look almost trivial.

The Lagrangian for a string against a background space is given as
\cite{string}
\bea
  L=\int e^{2\alpha}\Biggl(\frac{M_p^2}{2}(\partial_x \alpha)^2
\left( \mathcal{N}-\frac{{\mathcal N_1}^2}{
\mathcal{N}}\right)+M_p^2\frac{\partial_x \alpha\,
   \alpha^\prime {\mathcal N_1}}{\mathcal{N}}\nonumber\\-\frac{M_p^2\alpha^{\prime 2}
   {\mathcal N_1}\partial_x
  \bm \phi\,
    {\bm\phi} ^{\prime}}{2{\mathcal N}}+\frac{1}{2}(\partial_x {\bm \phi})^2 \left(\frac{ {\mathcal N_1}^2}{
   \mathcal{N}}-  \mathcal{N}\right)+\frac{ {\bm \phi} ^{\prime 2}}{2
   \mathcal{N}}\Biggr)d x,
   \label{ll}
\eea

\begin{figure}[ht]
\vspace{0. cm} \hspace{. cm}
 \includegraphics[width=8.5 cm]{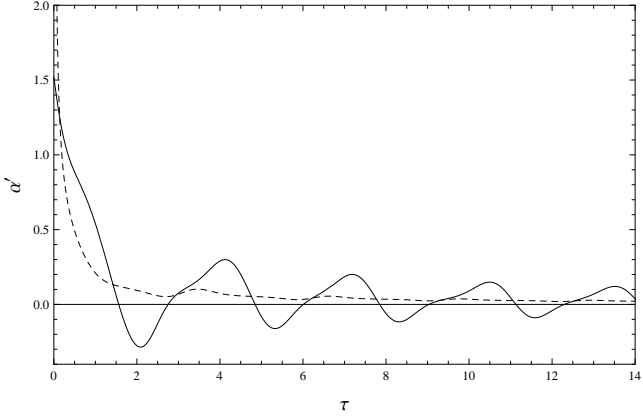}
 \vspace{0. cm}
 \caption{Evolution of  $\alpha^\prime(x,\tau)$ at $x=1.5$ according to Eq. (\ref{ev}). Numerical
 values of parameters are $\sqrt{N}/M_p=1.5$, $A=1.3$, $k=2$. Dashed line
 corresponds to $<\alpha^\prime>$, where the spatial averaging is implied.
    }
 \label{figs}
 \end{figure}

 \begin{figure}[ht]
\vspace{0. cm} \hspace{. cm}
 \includegraphics[width=8.5 cm]{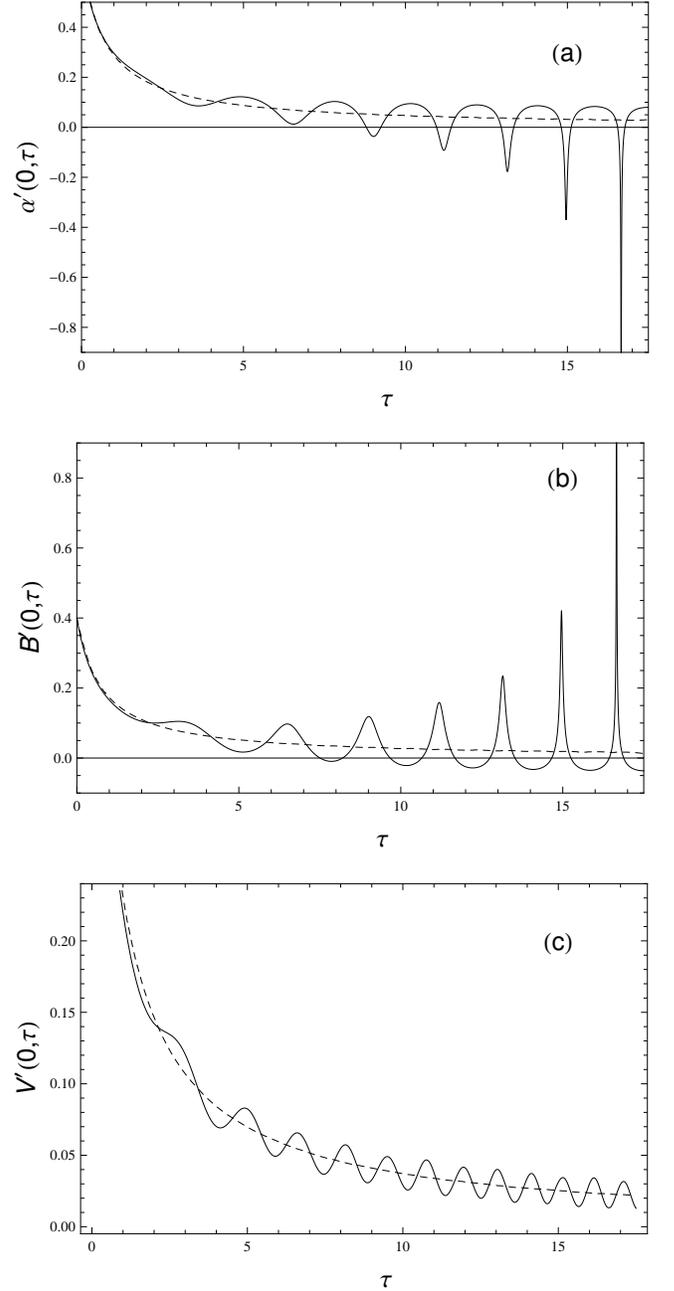}
 \vspace{0. cm}
 \caption{Evolution of
$\alpha^\prime(x,t)$ (a), $B^\prime(x,t)$ (b) and $V^\prime(x,t)$
(c) at $x=0$ for the inhomogeneous Bianchi I model. Numerical
values of the parameters are $\alpha_0 = -2$, $\pi_V = 1/100$,
$\pi_{B0} = 2/100$, $B_0 = 0$, $V_0(x)= \cos x/100 $. Dashed lines
correspond to the mean values $<\alpha^\prime>$, $<B^\prime>$ and
$<V^\prime>$.}
\label{fig1}
 \end{figure}

 \begin{figure}[h]
\vspace{0. cm} \hspace{. cm}
 \includegraphics[width=8.5 cm]{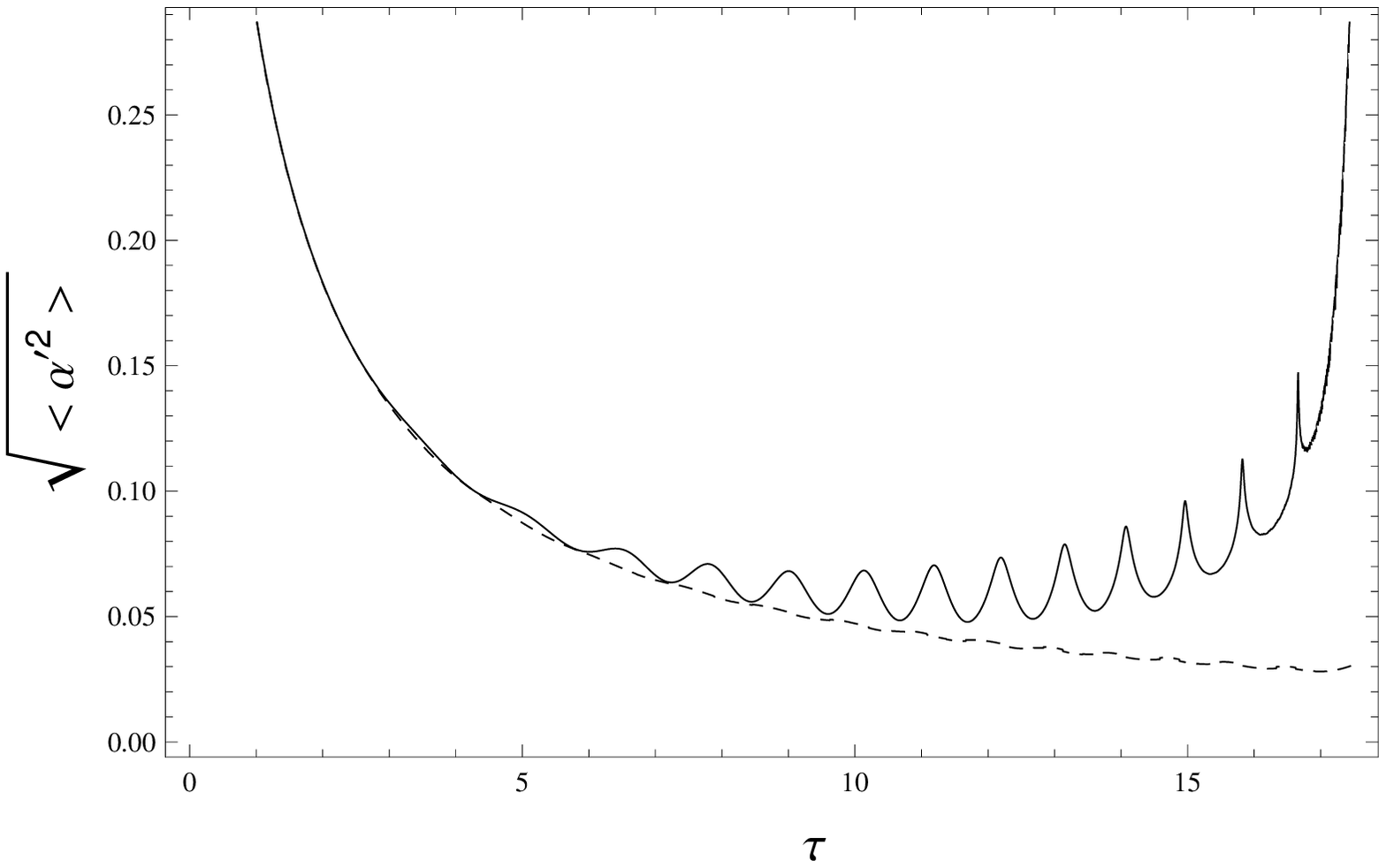}
 \vspace{0. cm}
 \caption{Evolution of the dispersion $\sqrt{<\alpha^{\prime 2}>}$. Dashed line
 corresponds to $<\alpha^\prime>$. }
 \label{fig2}
 \end{figure}

\noindent where $\bm \phi(x,\tau)=\{\phi_1(x,\tau),\phi_2(x,\tau)
\dots\phi_N(x,\tau) \}$ are the scalar fields, $\alpha(x,\tau)=\ln
a(x,\tau)$ is the logarithm of a ``scale factor" $a(x)$,
$\mathcal{N}$ and $\mathcal N_1$ are the lapse and shift
functions, respectively; $M_p$ is a constant (an analog of
the Plank mass); primes denote the derivatives over $\tau$. This
Lagrangian is not deduced directly from the general relativity
action, however, it can be reduced to the form corresponding to a flat
Friedman-Robertson-Walker (FRW) universe filled with the homogeneous
scalar fields (i.e. if both $\alpha$ and $\phi$ do
not depend on $x$). Thus Eq. (\ref{ll}) can be considered as a simplest
inhomogeneous generalization of the FRW Lagrangian.

Varying over $\mathcal N$ and $\mathcal N_1$ yields the
Hamiltonian and momentum constraints, which, in the case of $\mathcal
N=1$ and $\mathcal N_1=0$, take the form:
\bea  {\mathcal
H}=e^{2\alpha}\left(-M_p^2\alpha^{\prime
2}-M_p^2(\partial_x\alpha)^2+{\bm\phi}^{\prime
2}+(\partial_x{\bm\phi})^2\right)=0,\label{h1}\\
{\mathcal
P}=e^{2\alpha}\left(-M_p^2\alpha^\prime\,\partial_x\alpha
+{\bm\phi}^\prime\,\partial_x{\bm\phi}\right)=0.
\label{const}
\eea
Also, one can obtain the equations of motion \cite{string}
\bea
 \bm \phi^{\prime\prime}-\partial_{x x}\bm
\phi+2\alpha^\prime\bm \phi^\prime
-2\partial_{x}\alpha\partial_{x}\bm \phi=0,
~~~~~~~~~~~~~\label{e1}\\
M_p^2\alpha^{\prime\prime}-M_p^2\partial_{x  x
}\alpha+M_p^2\alpha^{\prime
2}-M_p^2(\partial_x\alpha)^2~~~~~~~~~\nonumber\\+\bm \phi^{\prime
2}-(\partial_x\bm \phi)^2=0.
\label{eqns1}
\eea
The constraints evolve according to the equations
\bea
 \mathcal H^\prime=\ptl_x \mathcal P,\nonumber\\
 \mathcal P^\prime=\ptl_x \mathcal H.
 \label{evconstr}
\eea
That is, if the constraints equal to zero initially, they remain to be zero
during all the evolution.

 Some particular solution  can be easily found
for the system (\ref{h1}), (\ref{const}), (\ref{e1}),
(\ref{eqns1}). Let us assume $\alpha(x,
\tau)=\frac{\sqrt{N}}{M_p}\phi_1(x,\tau)=\frac{\sqrt{N}}{M_p}\phi_2(x,\tau)=...\frac{\sqrt{N}}{M_p}\phi_N(x,\tau)$,
then the equations of motion (\ref{e1}),(\ref{eqns1}) are reduced
to the single equation
\be
\alpha^{\prime\prime}-\ptl_{xx}\alpha+\frac{2\sqrt{N}}{M_p}\left(\alpha^{\prime
2} -(\ptl_x\alpha)^2\right)=0.
\label{alf}
\ee
Also, one can be sure that the constraints (\ref{h1}), (\ref{const})
are satisfied, i.e. $\mathcal H=0$, $\mathcal P=0$.

It is interesting to note, that the analogous equation arises in
the the Szekeres theory of colliding waves \cite{sz}. However, the
initial conditions for this equation in the Szekeres theory are
restricted by the constraints, while in the present theory the
constraints are already satisfied and the initial conditions for
Eq. (\ref{alf}) are quite arbitrary.

The oscillations against a background can
appear if the initial values meet the
condition
\be
|\ptl _x\alpha(x,0)|>>|\alpha^\prime(x,0)|.
\ee

\noindent In the opposite case, a monotonic evolution dominates.

Solution of Eq. (\ref{alf}) can be written in the form of
\bea
 \alpha(x,\tau)=\frac{M_p}{2\sqrt{N}}\ln\biggl(\frac{1}{2}e^{\frac{2\sqrt{N}}{M_p}\alpha_0(x+\tau)}
+\frac{1}{2}e^{\frac{2\sqrt{N}}{M_p}\alpha_0(x-\tau)}\nonumber\\+\frac{\sqrt{N}}{M_p}\int_{x-\tau}^{x+t}k(\xi)d\xi\biggr),
\label{solal}
\eea
 \noindent where
the functions $\alpha_0(\xi)$ and $k(\xi)$ are connected with
the initial values of $\alpha$ and its derivative so that
$\alpha(x,0)=\alpha_0(x)$,
$\alpha^\prime(x,0)=e^{-2\frac{\sqrt{N}}{M_p}\alpha_0(x)}k(x)$.

As
an example, we take $k=const$, $\alpha_0(\xi)=A \cos\xi$. Then
the solution of Eq. (\ref{solal}) takes the form
\bea
\alpha(x,\tau)=\frac{M_p}{2\sqrt{N}}\ln\biggl(\frac{1}{2}\biggl(e^{2
A\cos(x-\tau)\sqrt{N}/M_p}~~~~~\nonumber\\+e^{2
A\cos(x+\tau)\sqrt{N}/M_p}\biggr)+\frac{2 k\sqrt{N}}{M_p}\tau
\biggr).
\label{ev}
\eea
Here, the global evolution is defined by the value of $k$ and
the oscillations are defined by the amplitude of $A$. The example of
evolution at some fixed coordinate $x$ is shown in Fig.
\ref{figs}.

From the point of view of the ``energy balance" specified by the
Hamiltonian constraint (\ref{h1}), the oscillations of scalar
fields do not affect a slow ``global" (i.e. ``background") evolution because they
are compensated by the oscillations of scale factor $\alpha$.

\section{Inhomogeneous Bianchi I model}
\label{secn}

As the next example, let analyse the inhomogeneous Bianchi I
model.  Hereinafter, we shall decompose the metric into the scale factor
$e^{2\alpha}$ and the components of conformal geometry, and use the conformal
time $\tau$.
 The argumentation for using the conformal time gauge is that it allows representing the equation of
motion for a scale factor of universe in a form
containing a difference of the potential and kinetic energies of
the field oscillators \cite{2,conf}.

Thus, the metric looks like
\bea
ds^2=e^{2\alpha}\biggl(d\tau^2-e^{-4 B}dx^2-e^{2
B+2\sqrt{3}\,V}dy^2\nonumber\\-e^{2 B-2\sqrt{3}\,V}dz^2\biggr),
\label{metric}
\eea
where the functions  $\alpha, B, V$ depend on $x$ and $\tau$.

The  Einstein equations allow obtaining the Hamiltonian and momentum
constraints as well as the equations of motion:

\begin{widetext}
\bea
  \mathcal H=\frac{1}{2}e^{2\alpha}\left(-\alpha^{\prime
2}+B^{\prime 2}+V^{\prime 2}\right)
 +e^{2\alpha+4 B}\left(\frac{1}{6}(\ptl_x
\alpha)^2+\frac{1}{3}\ptl_{xx}\alpha+\frac{7}{6}(\ptl_x
B)^2+\frac{1}{3}\ptl_{xx}B+\frac{4}{3}\ptl_x\alpha\ptl_x
B+\frac{1}{2}(\ptl_x V)^2\right),\label{hamcon}\\
  \mathcal P=
e^{2\alpha}\left(-\frac{1}{3}\ptl_x\alpha\alpha^\prime+\ptl_x B
B^\prime+\frac{2}{3}\ptl_x\alpha\, B^\prime+\ptl_x V
V^\prime+\frac{1}{3}\ptl_xB^\prime+\frac{1}{3}\ptl_x\alpha^\prime\right),~~~~~~~~
~~~~~~~~~~~~~~~~~~~~~~
~~~~~~~~~~~~~~~ \label{pconcon}\\
  \alpha^{\prime\prime}-e^{4B}\left(\ptl_{xx}\alpha
 +(\ptl_{x}\alpha)^2+(\ptl_x V)^2+
\frac{7}{3}(\ptl_x
B)^2+\frac{2}{3}\ptl_{xx}B+4\ptl_xB\ptl_x\alpha\right)+\alpha^{\prime
 2}+
V^{\prime
 2}+B^{\prime
 2}
 =0,~~~~~~~~~~~~~~~~~~~\label{11}\\
B^{\prime\prime}+\frac{1}{3}e^{4B}\left(\ptl_{xx}B+2(\ptl_{x}B)^2+6(\ptl_x
V)^2-2(\ptl_x\alpha)^2+2\ptl_x\alpha\ptl_xB+2\ptl_{xx}\alpha\right)
+2
B^\prime\alpha^\prime=0,~~~~~~~~~~~~~~\label{12}\\V^{\prime\prime}+2
V^\prime\alpha^\prime-e^{4B}\left(\ptl_{xx}
V+2\ptl_xV\ptl_x\alpha+4\ptl_xB\ptl_xV\right)=0.~~~~~~~~~~~~~~~~~~~~~~~
~~~~~~~~~~~~~~~~~~~~ \label{10}
\eea
\end{widetext}

It should be noted that the equations of motion can be also
obtained from the Hamiltonian $H=\int \mathcal H d x $ directly.
The constraint evolution is given by (\ref{evconstr}) as it takes place for a string.

Let ask: would the ripples of scale
factor appear if it was initially homogeneous. Since there is no analytical solution, we shall investigate a
particular solution of this model numerically.

 The initial
condition are taken in the form
\bea
\alpha(x,0)=\alpha_0=const,\nonumber\\
B(x,0)=B_0=const,\nonumber\\
V(x,0)=V_0(x)=A\cos(x),\nonumber\\
\eea
\bea
V^\prime(x,0)=\exp(-2\alpha_0)\pi_V(x),\nonumber\\
\alpha^\prime(x,0)=\exp(-2\alpha_0)\pi_\alpha(x),\nonumber\\
B^\prime(x,0)=\exp(-2\alpha_0)\pi_B(x),\nonumber\\
\eea
where the velocities are expressed in the terms of initial momenta
$\pi_V(x)$, $\pi_\alpha(x)$, $\pi_B(x)$, which are chosen as
\bea
\pi_V(x)=const,~~~~~~~~~~~~~~~~~~~~~~~~~~~~~~~~~~~~~~~~~ \nonumber\\
\pi_\alpha(x)=\frac{1}{2}(\pi_{B0}-3V_0(x)\pi_V~~~~~~~~~~~~~~~\nonumber\\
+\frac{\pi_V^2+e^{4B_0+4\alpha_0}V_0^{\prime2}(x)}{2(\pi_{B0}-3V_0(x)\pi_V)},\nonumber\\
\pi_B(x)=-\pi_\alpha(x)-3\pi_VV_0(x)+\pi_{B0},\nonumber\\
\pi_{B0}=const,~~~~~~~~~~~~~~~~~~~~~~
\eea
i.e. in such a way that the constraints (\ref{h1}), (\ref{const})
are satisfied exactly.

The evolution of time derivatives
 at $x=0$ is
shown in Fig. \ref{fig1}. One can see, that $\alpha^\prime(x,t)$
evolves non-monotonically and can become negative at some moment of time and in some
spatial point $x$, i.e. the scale factor decreases here. Analogous oscillations appear for $B^\prime$. Figure
\ref{fig2} shows the evolution of the dispersion of
$\alpha^\prime$, which finally becomes much greater than the mean
value $<\alpha^\prime>$.

\section{Linearized theory for the inhomogeneous Bianchi I model}

\label{seclin}

The origin of the oscillations demonstrated above can be easily explained within the framework
of a linearized theory. Let represent the variables as a sum of
a spatially uniform part and a perturbation
\bea
\alpha(x,\tau)=\alpha_0(\tau)+\alpha_1(\tau)e^{ikx},\nonumber\\
B(x,\tau)=B_0(\tau)+B_1(\tau)e^{ikx},\nonumber\\
V(x,\tau)=V_0(\tau)+V_1(\tau)e^{ikx}.\nonumber\\
\eea

In the zero-order in perturbations, the equations of motion are
\bea
{V_0}''+2 {V_0}' {\alpha_0}'=0,\nonumber\\
{B_0}''+2 {B_0}' {\alpha_0}'=0,\nonumber\\
{\alpha_0}''+{B_0}^{\prime2}+{V_0}^{\prime 2}+{\alpha_0}^{\prime
2}=0,
\label{zeroap}
\eea
and the only constraint  is
\be
\mathcal H_0=\frac{1}{2}e^{2\alpha_0}\left({B_0}^{\prime 2}+
{V_0}^{\prime 2}- {\alpha_0}^{\prime 2}\right). \label{h0}
\ee

\begin{figure}[ht]
\vspace{0.5 cm} \hspace{. cm}
 \includegraphics[width=8.5 cm]{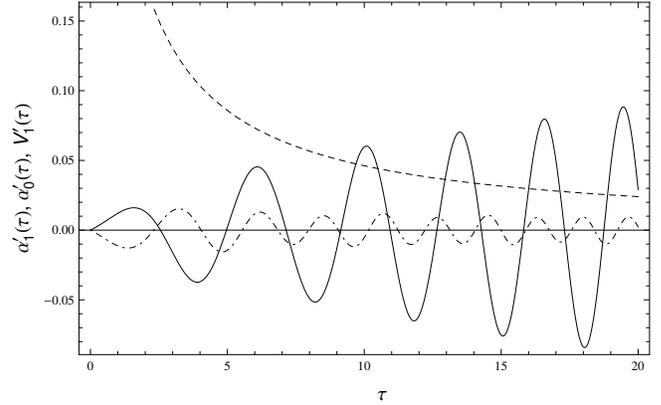}
 \vspace{0. cm}
 \caption{Evolution of $\alpha_0^\prime$ (dashed curve), oscillations of $\alpha_1^\prime$ (solid curve), and
$V_1^\prime$ (dotted curve) according to Eqs.
(\ref{zeroap}), (\ref{oneap}).
     }
 \label{figlin}
 \end{figure}

\begin{figure*}
\vspace{0. cm} \hspace{. cm}
 \includegraphics[width=17.5 cm]{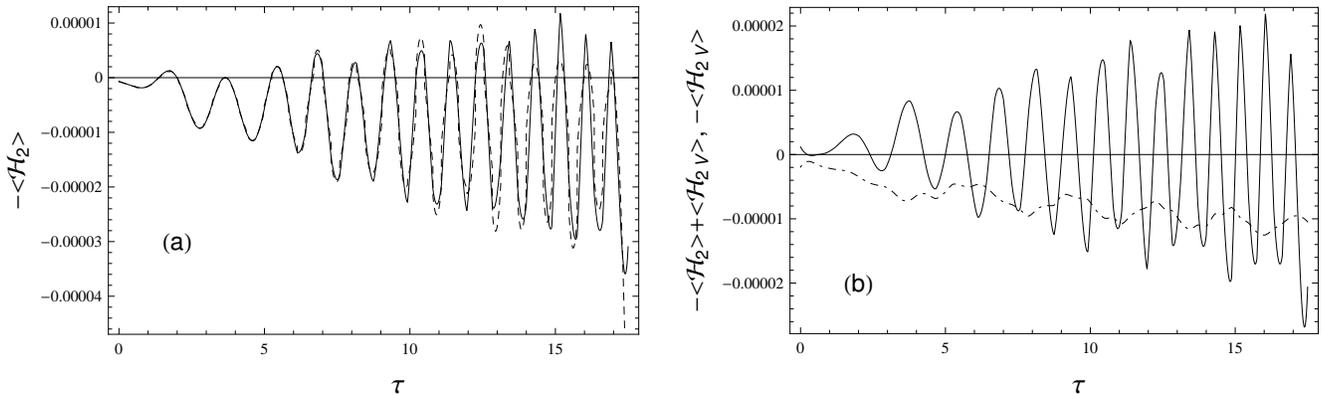}
 \vspace{. cm}
 \caption{ (a) Solid line is the mean value of  the second-order contribution
of $\xi_1$  to the Hamiltonian constraint. Dashed line is the
background evolution i.e. $\mathcal H_0$ given by (\ref{h0}).
 (b) Solid line is the mean value of the second-order contribution, but without taking into account a ``pure"
 gravitational wave $<\mathcal H_{2V}>$. Dashed line is
 the contribution of a pure gravitational wave, namely $<\mathcal  H_{2V}>$.
} \label{fig3}
 \end{figure*}

In the first-order in perturbations, one has
\begin{widetext}
\bea
 {V_1}''+k^2 e^{4 {B_0}} {V_1}+2
{V_0}' {\alpha_1}'+2 {V_1}' {\alpha_0}'=0,\nonumber\\
   {B_1}''+2{B_0}' {\alpha_1}'+2
{B_1}'
   {\alpha_0}'- \frac{1}{3}k^2 e^{4 {B_0}} ({B_1}+2
{\alpha_1})=0,\nonumber\\
  {\alpha_1}''+2
   {\alpha_0}' {\alpha_1}'+2 {B_0}' {B_1}'+2
{V_0}' {V_1}'
   +\frac{1}{3}
k^2 e^{4 {B_0}} (2 {B_1}+3 {\alpha_1})=0,\nonumber\\
  \mathcal H_1=-3 {B_0}'{B1}'-3 {\alpha_1}
{B_0}^{\prime 2}+k^2 e^{4 {B_0}} {B_1}+k^2 e^{4 {B_0}}
   {\alpha_1} -3 {V_0}' {V_1}'-3 {\alpha_1} {V_0}^{\prime 2}\nonumber\\
   +3 {\alpha_0}' {\alpha_1}'+3 {\alpha_1} {\alpha_0}^{\prime 2}=0,\nonumber\\
     \mathcal P_1=-3 i k {B_1} {B_0}'-2 i k {\alpha_1} {B_0}'
  -i k {B_1}'-3 i k {V_1} {V_0}'+i k
   {\alpha_1} {\alpha_0}'-i k {\alpha_1}'=0.
\eea
\end{widetext}
However, only one of the two last constraints turns out to be
independent. Thus one may exclude $B_1$ and after some computations
comes to the equations  for $\alpha_1, V_1$, which can be solved
numerically:
\begin{widetext}
\bea
{V_1}''+k^2 e^{4 {B_0}} {V_1}+2 {V_0}' {\alpha_1}'+2 {V_1}'
{\alpha_0}'=0,~~~~~~~~~~~~~~~~~~~~~~~~~~~~~~~~~~~~~~~~~~~~~
~~~~~~~~~~~~~~~~~\nonumber\\
 {\alpha_1}''-\frac{18 {B_0}' {B_0}'' {\alpha_1}'+\frac{1}{3}e^{8 {B_0}} k^4 {\alpha_1}\
 + e^{4 {B_0}} k^2 \left(4 {V_0}' \left({V_1}'-3
{V_1} {B_0}'\right)-4 {B_0}' {\alpha_1}'+{\alpha_1} \left(7
   \left({B_0}'\right)^2-2 {B_0}''\right)\right)}{{ 9
   \left({B_0}'\right)^2+e^{4 {B_0}} k^2}}=0.
   \label{oneap}
\eea
\end{widetext}
 Evolution of $\alpha_1^\prime(\tau)$,
$V_1^\prime(\tau)$ is shown in Fig. \ref{figlin}, which demonstrates that the amplitude of oscillations of $\alpha_1^\prime(\tau)$
increases with time. For the primary system (\ref{hamcon}),(\ref{pconcon}), (\ref{11}),
(\ref{12}), (\ref{10}), the oscillations (Fig. \ref{figlin})
transform into the narrow spikes due to nonlinearity (see Fig. \ref{fig1}(a)).

\section{Ripples from the point of view of the Isaacson's theory}
\label{secis}

The Isaacson's theory \cite{isa} considers an influence of the ripples on
the evolution of background. An interpretation of the Isaacson's theory is
quite straightforward and does not depend on details of the
averaging procedure. Let expound its main aspects.  Let suppose, that
there is a set of equations
\be\bm
A(\bm \xi(x,\tau))=0 \label{eqA}
\ee
 for some  quantities $\bm \xi$, which have a
strongly inhomogeneous (oscillating) space-time behavior. Then, one
may define an average quantity
\be
\bm \xi_0(\tau)=<\bm \xi(\tau,x)>=\frac{1}{L}\int_0^L\bm
\xi(\tau,x)dx,
\ee
where we shall consider only spatial averaging for the sake of simplicity.

As a result, the quantity $\bm \xi$ can be separated as
\be
\bm \xi(x,\tau)=\bm \xi_0(\tau)+\bm \xi_1(x,\tau),
\ee
where $\bm \xi_1(x,\tau)=\bm \xi(x,\tau)-\bm \xi_0(\tau)$. It is
evident, that the mean value of $\bm \xi_1(x,\tau)$ equals to zero.

Eqs. (\ref{eqA}) can be expanded in powers of $\bm
\xi_1(x,t)$:
\bea
  \bm A(\bm \xi(x,\tau))=\bm A_0(\bm \xi_0(\tau))+\bm A_1(\bm
\xi_0(\tau),\bm \xi_1(x,\tau))\nonumber\\+\bm A_2(\bm
\xi_0(\tau),\bm \xi_1(x,\tau))+\dots=0.
\label{eqAexp}
\eea
The first statement of the theory considered reads that the spatially
inhomogeneous part obeys the linear equations $\bm A_1=0$, and, in fact,
is equivalent to the linearization considered above. This
statement is not valid in our case, because the
ripples of $B$ and $\alpha$ are strongly nonlinear.

The second statement reads that the evolution of
background is determined by averaged terms of the second
order in ripples:
\be
\bm A_0(\bm \xi_0(\tau))=-<\bm A_2(\bm \xi_0(\tau),\bm
\xi_1(x,\tau))>.
\ee

It is a consequence of the averaging of Eq.
(\ref{eqAexp}).

Let consider only single equation: the Hamiltonian
constraint (\ref{hamcon}). We substitute
\bea
\alpha(x,\tau)=\alpha_0(\tau)+ \alpha_1(\tau,x),\nonumber\\
B(x,\tau)=B_0(\tau)+B_1(\tau,x),\nonumber\\
V(x,\tau)=V_0(\tau)+V_1(\tau,x)\nonumber\\
\eea
into the Hamiltonian constraint and obtain the second order
contribution in the form of
\begin{widetext}
\bea
\mathcal H_{2}=e^{2\alpha_0}\biggl(2  {\alpha_1}  {\ptl_x B_1}
 {B_0}'+e^{4  {B_0}}\biggl(\frac{4}{3}  {\ptl_x B_1}  {\ptl_x \alpha_1}
+\frac{2}{3}  {\alpha_1}
    {\ptl_{xx}B_1} +\frac{4}{3}  {B_1}  {\ptl_{xx}B_1}+\frac{7}{6}
   \left( {\ptl_x B_1}\right)^2+\frac{4}{3}   {B_1}  {\ptl_{xx}\alpha_1}\nonumber\\+\frac{1}{2}
   \left( {\ptl_xV_1}\right)^2+\frac{1}{6} \left( {\ptl_x \alpha_1}\right)^2 +\frac{2}{3}  {\alpha_1}
    {\ptl_{xx}\alpha_1}\biggr)+\frac{1}{2} \left( {\ptl_x B_1}\right)^2+2  {\alpha_1}  {\ptl_x V_1}  {V_0}'+\frac{1}{2}
   \left( {V_1}^{\prime}\right)^2-2  {\alpha_1}  {\alpha_1}^{\prime}  {\alpha_0}'\nonumber\\-\frac{1}{2} \left(
   {\alpha_1
   }^{\prime}\right)^2+ {\alpha_1}^2 \left(\left( {B_0}'\right)^2+  \left( {V_0}'\right)^2-  \left(
   {\alpha_0}'\right)^2\right)\biggr).
\eea
\end{widetext}
Using the results of the previous numerical calculations (Sec.
\ref{secn}), one can see that the second statement of the
Isaacson's theory (i.e. $\mathcal H_0\approx-<\mathcal H_2> $) is
satisfied with a high accuracy as it is shown in Fig .
\ref{fig3} (a) .

Let distinguish a ``pure" gravitational wave
contribution
\be
\mathcal H_{2V}=e^{2\alpha_0}\frac{1}{2}\left(\left(
{V_1}^{\prime}\right)^2+\left( {\ptl_xV_1}\right)^2\right)
\ee
in $\mathcal H_2$.  As one can see from Fig. \ref{fig3} (b),
the plot of  $<\mathcal H_2-\mathcal H_{2V}>$ is, in fact, symmetric
relatively the time axis. This means that if one performs the time
averaging in addition, then one can
conclude that $<\mathcal H_{2}-\mathcal H_{V2}>$ does not
contribute ``at the mean" to the evolution of background, i.e.
the evolution of background is determined by a ``pure"
gravitational wave $<\mathcal H_{2V}>$.

\section{The York curvature for the nonuniform Bianchi model}
Why is the $V$-variable considered as that corresponding to a ``pure"
gravitational wave? To answer this question, one may use the description of a conformal three-geometry in the terms of the York
curvature \cite{York, mizn}
\be
Y^{ab}=g^{1/3}e^{aef}(R^{b}_f-\frac{1}{4}\delta^b_f R)_{;e},
\ee
where $g_{ab}$ denotes the spatial part of the metric (\ref{metric}),
$R_{ab}$ is the Ricci tensor of the three-metric $g_{ab}$, $e^{aef}$ is the
fully antisymmetric symbol and the ${;e}$ means the covariant derivative
over the metric $g_{ab}$. Direct calculation gives the following
result
\bea
Y^{23}=\sqrt{3} e^{4 B} \biggl(9 \ptl_x B \ptl_{xx}V+3 \ptl_xV
\biggl(\ptl_{xx}B+6 (\ptl_x B)^2\nonumber\\
-2 (\ptl_xV)^2\biggr)+\ptl_{xxx}V\biggr).~~~
\label{York}
\eea
Other components of $Y^{ab}$ equal to zero. As can expect,
the York curvature is conformally invariant, i.e. it does not
depend on $\alpha$. From Eq. (\ref{York}), one can see
that $V=0$ results in $Y=0$ and the space becomes conformally flat.
This justifies the consideration of the $V$-variable as representing
a ``pure" gravitational wave.

\section{Discussion and conclusions}

Let compare the model considered with the well-known Gowdy
\cite{gowdy,mizng,berger,berger1} model described by the metric:
\be
ds^2=e^{- T-\lambda/2}\left(e^{4 T}d t^2-dx^2\right)-e^{2
T}(e^{2\beta}dy^2+e^{-2\beta}dz^2),
\ee
where $T,\lambda,\beta$ are the functions of $t,x$ only. Denoting
\bea
\beta=\sqrt{3}V,\nonumber\\
T=B+\alpha,\nonumber\\
\lambda=6(B-\alpha),
\eea
we come to the metric
\bea
ds^2=e^{2\alpha}\bigl((e^{2\alpha}dt)^2-e^{-4 B}dx^2-e^{2
B+2\sqrt{3}\,V}dy^2\nonumber\\-e^{2 B-2\sqrt{3}\,V}dz^2\bigr),
\eea
which is exactly the metric (\ref{metric}) if the conformal
time $d\tau=e^{2\alpha}dt$ is used instead of $t$.

For the Gowdy model, it is shown that the function $T$ can be chosen as the
function of the time $t$ only \cite{gowdy,berger,berger1}, and the function $\lambda$ ceases
to depend on $x$ at $t\rightarrow \infty$. Thus, in principle, the ripples of
$\alpha=\frac{1}{2}(T-\frac{\lambda}{6})$ found above can
be eliminated by some coordinate transformation.

It should be noted, that, if the gauge invariance turns out to be
violated in quantum theory, the ripples of the scale factor in the
conformal time can appear. Since, as it has been demonstrated
above (Figs. \ref{fig2}, \ref{fig3}), the ripples consist of the
sharp spikes, their spectrum will be broad and ``noise-like". It
should be taken into account, that such spikes have appeared when
the Bianchi I model was considered. As has been demonstrated in Sec.
\ref{seclin}, the unstable rising modes exist for this model. In
contrast, it is known, that the isotropic FRW space-time is stable
against small perturbations \cite{lan}. Thus, for the isotropic
space-time, the ripples could exist only in a strongly nonlinear
regime and this issue needs a further analysis.

Concerning a contribution of the ripples to the energy balance, one may conclude that
such ripples occur in both string and Bianchi I models. But, in the string model the ripples of the scale factor and the oscillations of the matter fields compensate mutually euch other and, thereby, the background evolution is not affected by them. For the Bianchi I
model, the results of Sec. \ref{secis} demonstrate that the
ripples of the scale factor compensate only the oscillations of the $B$-field
and the background evolution ``at the mean" becomes to be determined by the
``pure" gravitational wave $V$ in agreement with the results of
\cite{bon}. Nevertheless, such a conclusion can become invalid in
the quantum gravity case, since a strong gauge wave,
consisting of the scale factor $\alpha$ and the $B$-field could
mixed up the gravitational waves and contribute to the background evolution, as well.

It should be noted, that both models (i.e. the string and Bianchi I
ones) have been considered on the classical level that cannot shed a light on the solution of the vacuum energy
problem. This results from the fact that a compensation of only
zero-point fluctuations of matter fields is required in the quantum case. But in the classics,
it is impossible to distinguish the fluctuations in the ground state from those in an excited state.

\addcontentsline{toc}{section}{References}

\end{document}